%% file: prog_comput_4.tex
\documentclass[twocolumn,aps,pra,showpacs,groupedaddress]{revtex4}
\usepackage{amssymb,mathrsfs}
\usepackage[active]{srcltx}
\input{myQcircuit}

\def\<{\langle}\def\>{\rangle}\def\map#1{\mathcal #1}\def\Uset{{\mathscr U}}
\def\QS{\texttt{QS}}\def\CNOT{\texttt{CNOT}}
\newtheorem{theorem}{Theorem}\newtheorem{task}{Task}
\def\qed{$\,\blacksquare$\par}\def\Proof{\par\noindent{\bf Proof. }}
\begin{document}
\title{Quantum computation with programmable connections between
  gates} \author{Timoteo Colnaghi}\email{timoteo.colnaghi@gmail.com}
\affiliation{{\em QUIT Group}, Dipartimento di Fisica,
  via Bassi 6, 27100 Pavia, Italy} \author{Giacomo Mauro
  D'Ariano}\email{dariano@unipv.it} \affiliation{{\em QUIT Group},
  Dipartimento di Fisica, via Bassi 6, 27100 Pavia,
  Italy} \affiliation{Istituto Nazionale di Fisica Nucleare, Gruppo
  IV, Sezione di Pavia, Italy} \homepage{http://www.qubit.it}
\author{Paolo Perinotti}\email{paolo.perinotti@unipv.it}
\affiliation{{\em QUIT Group}, Dipartimento di Fisica,
  via Bassi 6, 27100 Pavia, Italy} \affiliation{Istituto Nazionale di
  Fisica Nucleare, Gruppo IV, Sezione di Pavia, Italy}
\homepage{http://www.qubit.it} \author{Stefano
  Facchini}\email{stefano.facchini@unipv.it} \affiliation{Research
  Center for Quantum Information, Slovak Academy of Sciences,
  D\'ubravsk\'a cesta 9, 845 11 Bratislava, Slovakia}
\homepage{http://www.quniverse.sk} 
\date{\today}
\begin{abstract}
  A new model of quantum computation is considered, in which the connections between gates are
  programmed by the state of a quantum register. This new model of computation is shown to be more
  powerful than the usual quantum computation, e.~g. in achieving the programmability of
  permutations of $N$ different unitary channels with $1$ use instead of $N$ uses per channel.
  For this task, a new elemental resource is needed, the {\em quantum switch}, which can be programmed to
  switch the order of two channels with a single use of each one.
\end{abstract}
\maketitle
Quantum computation \cite{ncbook} has revolutionized computer science, showing that the processing
of quantum states can lead to a tremendous speedup in the solution of a class of problems, as
compared to traditional algorithms that process classical bits. On the other hand, in classical
computation it is customary to design algorithms that treat subroutines in exactly the same way as
data---the {\em program as data} paradigm inspired by the Church's notion of computation
\cite{barend}---which allows one to compute functions of functions, rather than only functions of
bits. Such a programming strategy, however, is not an option in the quantum case. This is due to the
fact that in quantum computation data and subroutines are intrinsically different objects---the
former being quantum states, the latter unitary transformations---and exact programming of unitary
transformations over quantum states is impossible with finite resources \cite{niels}.

A new kind of computational model is needed for a quantum functional calculus that uses unitary
transformations (generally channels) as input and output subroutines.  This paradigm has been
recently established and developed theoretically, based on the notion of {\em quantum combs and
  supermaps} \cite{qca,comblong}, which describe circuit-boards in which quantum channels can be
plugged. Thus, quantum combs describe any kind of coherent adaptive quantum strategy \cite{watgut},
e.~g. any quantum algorithm calling oracles (the oracles are the plugged-in channels).  Coherent
adaptive strategies are the most general architecture allowed in the quantum circuit model and have
been demonstrated to be more powerful than non-adaptive ones for the problem of channel
discrimination \cite{memorydisc,Harrow}.  They are also very useful for quantum computation, as many
quantum algorithms consist in the discrimination of classes of oracles.

There exist tasks, however, that even a coherent adaptive quantum strategy cannot accomplish: for
example, an algorithm that takes two unknown boxes and, using the boxes a {\em single time},
connects them in the two possible orders depending on the value of a control qubit. Even more: the
algorithm produces a quantum superposition of the two orderings. As shown in Ref.
\cite{beyondqc}, realizing such a ``quantum switch'' within the standard circuit model would
unavoidably need {\em two uses} of each box.  More generally, if we want to achieve the
superposition of all possible permutations of the orderings of $N$ boxes, many uses per box will be
needed (here it will be shown that the minimum number of uses is $N$).  On the other hand, as
already outlined in Ref.  \cite{beyondqc} for $N=2$, by using a coherent switching of paths, e.g. in
the optical domain, one would need only a single use per oracle.

The number of uses of a box on computational grounds represents the so-called {\em query
  complexity}. This is a primary resource to be taken into account, and in some cases  it actually corresponds
to the number of physical copies of the device that are needed.  A coherent switching of
input-output connections can have a dramatic impact on the performances of quantum computers,
since the use of superpositions of permutations of oracle-calls can be of enormous advantage for oracle
discrimination, thus achieving e.g. a fast database search with a low-complexity optical setup.

In this letter we consider such a new kind of quantum computation, where the connections
between gates are themselves programmable on the state of quantum registers. We will show that
these dynamical quantum networks are more powerful than the the usual quantum computers, in the sense that they reduce the query complexity for some tasks. We will prove
this for the case of programming the permutations of $N$ different unitary channels, where
the number of uses of each input channel is dramatically reduced from $N$ to $1$.    For this task a new
resource is needed, the {\em quantum switch} (abbreviated as \QS), which can programmed to switch
the order of two channels with a single use per channel. We will represent the action of the \QS\
with the following graph
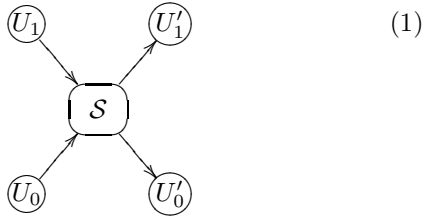
\begin{equation}
\xymatrix @C=.5em @R=1em @! {
*+[o][F-]{U_1} \ar[dr] & & *+[o][F-]{U'_1}  \\
 & *+<12pt>[F-:<6pt>] {\,\map S\, } \ar[ur] \ar[dr] & \\
*+[o][F-]{U_0} \ar[ur] & & *+[o][F-]{U'_0}
}
\end{equation}
where $\map{S}$ represents the \QS\ having two inputs and two outputs, $U_0$ and $U_1$ denote the
two input unitary channels, and $U'_0$ and $U'_1$ are the two output ones, which are either
$U'_0=U_0$ and $U'_1=U_1$ or $U'_0=U_1$ and $U'_1=U_0$, depending on the state $|0\>$ or $|1\>$ of a
control qubit $s$ (not represented). In formula

\begin{equation}
\map S(U_1\otimes U_0\otimes|i\>)=U_{1\oplus i}\otimes U_{0\oplus i}\otimes|i\>,
\end{equation}
where the position of $U_j$, $j=0,1$ in the tensor product determines the order in which $U_j$ is
applied in the circuit in which the \QS\ is inserted (channel $U_{0\oplus i}$ first). Notice
that in principle, other fixed operations can be performed between $U'_0$ and $U'_1$. 

In order to compare the power of customary quantum computation within the circuit model with that of dynamically
programmable quantum networks, we compare their performances in the task of {\em controlled
  permutation}, namely the task of programming all possible permutations of $N$ unitary channels $\{U_i\}_{i
  = 0}^{N-1}$ acting in cascade on one target qubit. First we design the most efficient conventional
quantum network that achieves controlled permutation, and show that $N$ uses per channel are needed.
As we will see the optimal network needs $\mathcal{O}(N^2)$ \CNOT s and $\mathcal{O}(N^2)$ control
qubits.  We then build a network made only of quantum switches that can program all the permutations
of the $N$ unitary channels with a single use per channel, and then we prove optimality of this
network in minimizing the number of \QS s and of control qubits.

\medskip
Consider the two following tasks:
\begin{task} \label{t1} Let $\Uset := \{U_i \}_{i=0}^{N-1}$ be a set of $N$ different unitary
  single-qubit operators. Build an efficient quantum circuit that lets a (unknown) qubit state $\ket{\psi}$
  undergo one of the $N^N$ dispositions of the unitaries in $\Uset$, with the specific disposition
  programmed on the state of a control register.
\end{task}
\begin{task} \label{t2} Build an efficient quantum circuit that outputs the state $U_{i}\ket{\psi}$
  for any $U_i\in\Uset$, with $i$ programmed on the state of a control register.
\end{task}
In the following we will show that Task \ref{t1} has a straightforward solution in terms of Task
\ref{t2}. For this reason, we first solve Task \ref{t2}. We restrict to $N$ power of $2$, the case
of general $N$ following straightforwardly. For $i\in[0,2^n]$ integer we denote by $[i]_n\in\{ 0, 1
\}^{n}$ the binary $n$-string representation of $i$.

Consider the following circuit composed of two registers, a {\em system} and a {\em control}, made
of $N$ qubits and $n=\log N$ qubits respectively:
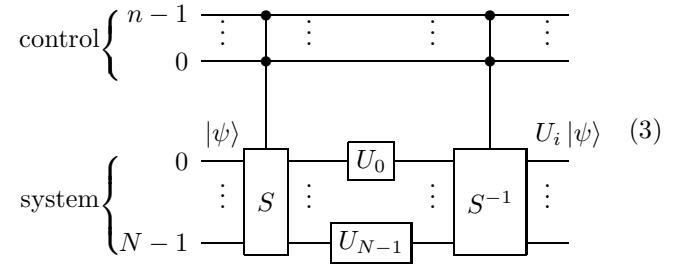
\begin{equation}
\begin{aligned}\label{CnSWAPgenerico}
\Qcircuit @C=0.8em @R=.5em { 
&&& & \lstick{n-1} &   \qw & \ctrl{7} & \qw  & \qw & \qw &  \ctrl{7} & \qw & \qw \\
\lstick{\text{control\,}} &&& &  & \vdots & & \vdots &  & \vdots & & \vdots &  \\
&&& & \lstick{0}  & \qw & \ctrl{5} & \qw & \qw & \qw &  \ctrl{5} & \qw & \qw \\
&&& & & & & & & & & &\\
&&& & & & & & & & & &\\
&&& & & & & & & & & & \\
&&& & & & & & & & & &\\
&&& & \lstick{0}  &  \ustick{\ket{\psi}} \qw & \multigate{2}{S} & \qw   & \gate{U_{0}} & \qw & \multigate{2}{S^{-1}} & \qw & \ustick{U_{i} \ket{\psi}} \qw \\ 
\lstick{\text{system\,}} &&&  &  & \vdots &  & \vdots  &  & \vdots & & \vdots & \\ 
&&& & \lstick{N-1} & \qw & \ghost{S} & \qw & \gate{U_{N-1}} & \qw & \ghost{S^{-1}} & \qw &  \qw
\gategroup{1}{1}{3}{1}{.5em}{\{}
\gategroup{8}{1}{10}{1}{.5em}{\{}
} 
\end{aligned}
\end{equation}
In the circuit, the system register is prepared in the state $\ket{\psi}\otimes \ket{0}^{N-1}$ and
the control in the state $|[i]_n\>$. This will output the state $U_i|\psi\>$ on the first system
qubit. For this purpose one first operates an $n$-controlled swap $S$, then each system qubit $i$
undergoes $U_i$, and finally an inverse $n$-controlled swap $S^{-1}$ is made. The $n$-controlled
swap is achieved by the following circuit:
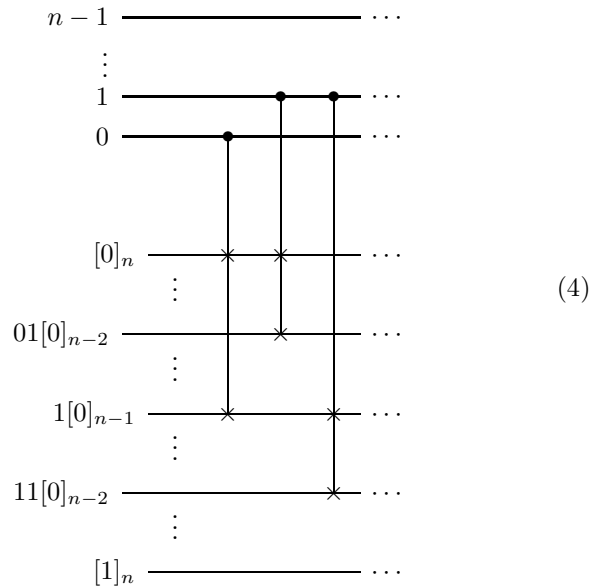
\begin{equation}\label{CnSWAP}
\begin{aligned}
\Qcircuit @C=1em @R=1.5em { 
 \lstick{n-1} & \qw & \qw  & \qw & \qw  & \qw & \qw & \qw & \qw & \qw & \cdots & \\   %
\lstick{\vdots} & \\
 \lstick{1} & \qw & \qw & \qw & \qw & \qw & \ctrl{6} & \qw & \ctrl{10} & \qw  & \cdots & \\
  \lstick{0} & \qw & \qw & \qw & \ctrl{5} & \qw & \qw & \qw & \qw & \qw & \cdots & \\
 & \\   %
 &  \\  %
 &  \lstick{[0]_n}  & \qw & \qw & \qswap & \qw & \qswap & \qw & \qw & \qw & \cdots & \\
  & & \vdots &\\
  \lstick{ 01[0]_{n-2}} & \qw & \qw & \qw & \qw & \qw & \qswap & \qw & \qw & \qw & \cdots & \\
 & & \vdots & & \qwx & \\
 &  \lstick{1[0]_{n-1}} & \qw & \qw &  \qswap \qwx & \qw & \qw & \qw & \qswap & \qw & \cdots &\\
 & & \vdots & \\
 \lstick{11[0]_{n-2}} & \qw & \qw & \qw & \qw & \qw & \qw & \qw & \qswap & \qw & \cdots & \\
 &  & \vdots & \\  %
  & \lstick{[1]_n}  & \qw & \qw & \qw & \qw & \qw & \qw & \qw & \qw & \cdots & \\  %
} 
\end{aligned}
\end{equation}
In words, for $k=0$ to $n-1$ the circuits does the following: insert $2^k$ control-swaps controlled
by the $(k+1)$-th control qubit and swapping system qubit $[s]_k[0]_{n-k-1}$ with system qubit
$[s]_k1[0]_{n-k-2}$ for every $k$-bit string $[s]_k$. 

\begin{theorem} \label{circT2optimality} Circuit (\ref{CnSWAP}) achieves Task \ref{t2} in the most
  efficient way
\end{theorem}

\Proof The minimum number of bits needed to write an integer smaller than $N$ is $n=\log N$, whence
the number of controlling qubits is optimal. Then, the best algorithm to single out an element $U_i$
out of the set $\Uset$ of $N$ unitaries is given by a sequence of bisections of $\Uset$, seeking at
each step to which half $U_i$ belongs, and encoding $0$ or $1$ accordingly, which is exactly the
strategy of Circuit (\ref{CnSWAP}), whence also the number of control-swaps is minimal.\qed

We need $2^{k}$ control-swaps to implement the $k$th steps, whence we need $\sum_{i=0}^{n-1}2^k =
2^n -1=N -1$ control-swaps in total to perform the $n$-controlled swap. Knowing that a control-swap
can be achieved by the sequence of a \CNOT, a Toffoli, and a \CNOT, and that the optimal
implementation of the Toffoli gate requires 6 \CNOT s and 9 single qubit operations
\cite{SHENDEMARKOV}, we conclude that $8(N-1)$ \CNOT s and $9(N-1)$ single-qubit operations are
needed to achieve a $n$-controlled swap, whence $16(N-1)$ \CNOT s and $18(N-1)$ single-qubit
operations are necessary to achieve Task \ref{t2}.  As an example, for $N = 8$ the $3$-controlled
swap is achieved with 7 control-swaps as follows
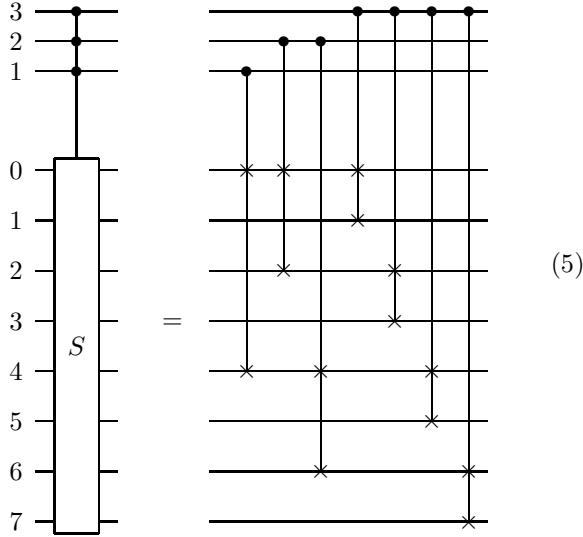
\begin{equation}
\begin{aligned}\label{CnSWAP8}
\Qcircuit @C=.7em @R=1em { 
 \lstick{3} & \ctrl{1}& \qw && &&& & \qw & \qw& \qw & \qw & \qw & \qw& \qw & \ctrl{5} & \qw & \ctrl{7} & \qw & \ctrl{9} & \qw & \ctrl{11} & \qw \\   %
 \lstick{2} & \ctrl{1}& \qw&& &&& & \qw & \qw & \qw & \ctrl{4} & \qw & \ctrl{8} & \qw & \qw & \qw & \qw & \qw & \qw & \qw & \qw & \qw \\   %
 \lstick{1} & \ctrl{3}& \qw&& &&&  & \qw & \ctrl{3} & \qw & \qw &\qw & \qw & \qw & \qw & \qw & \qw & \qw & \qw & \qw & \qw & \qw \\  %
 && && & & & & & & & && & & & & & & & &\\   %
 &&&& & & & & & & & & & && & & & & & & \\  %
 \lstick{0} &\multigate{7}{S}&\qw&& & && & \qw & \qswap & \qw &\qswap  &\qw &\qw &\qw & \qswap & \qw & \qw & \qw & \qw &\qw & \qw & \qw \\   %
\lstick{1} &\ghost{S} &\qw& & &&  &   & \qw & \qw \qwx & \qw  &\qw \qwx &\qw &\qw &\qw & \qswap \qwx & \qw & \qw & \qw & \qw &\qw & \qw & \qw \\  %
 \lstick{2}&\ghost{S}&\qw& && && & \qw & \qw \qwx & \qw  &\qswap \qwx &\qw &\qw & \qw & \qw & \qw & \qswap  & \qw & \qw &\qw & \qw & \qw \\  %
\lstick{3} &\ghost{S}&\qw& && = && & \qw & \qw \qwx & \qw  &\qw &\qw &\qw & \qw & \qw  & \qw & \qswap \qwx & \qw & \qw &\qw & \qw & \qw \\  %
 \lstick{4}& \ghost{S}&\qw&&& && & \qw & \qswap \qwx & \qw  &\qw &\qw &\qswap &\qw & \qw & \qw & \qw & \qw & \qswap & \qw &\qw & \qw  \\  %
\lstick{5} &\ghost{S}&\qw& && && & \qw & \qw & \qw &\qw &\qw &\qw \qwx  &\qw & \qw & \qw & \qw & \qw & \qswap \qwx & \qw &\qw & \qw  \\  %
\lstick{6} &\ghost{S}&\qw& && && & \qw & \qw & \qw  &\qw &\qw &\qswap \qwx &\qw & \qw & \qw & \qw & \qw & \qw & \qw &\qswap & \qw \\  %
\lstick{7} &\ghost{S}&\qw& && && & \qw & \qw & \qw &\qw &\qw &\qw &\qw & \qw & \qw & \qw & \qw & \qw & \qw &\qswap \qwx & \qw  \\  %
} 
\end{aligned}
\end{equation}

The result of theorem \ref{circT2optimality} allows us to conclude that Task \ref{t1} is achievable
by an ordinary quantum circuit on $\mathbf{O}(N \log{N})$ ancillary qubits plus $1$ system qubit,
and resorting to $\mathbf{O}(N^{2})$ elementary operations, as follows
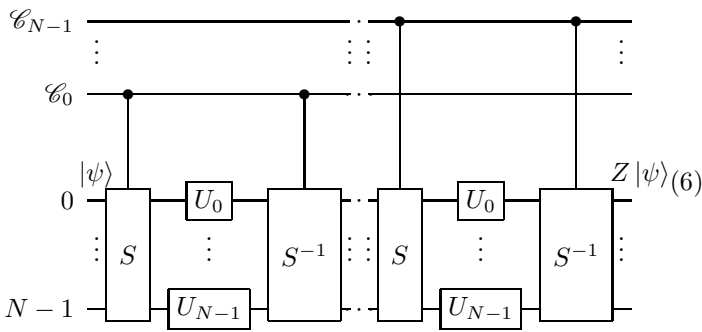
\begin{equation} \label{circT1}
\begin{aligned}
\Qcircuit @C=.35em @R=1em { 
\lstick{\mathscr{C}_{N-1}} & \qw & \qw & \qw & \qw & \qw & \qw & \qw & \cdots &   & \ctrl{5} & \qw & \qw & \qw & \ctrl{5} & \qw & \qw\\
& \vdots &   &   &   &   &   & \vdots & & \vdots &   &   &   &   &   & \vdots  & \\
\lstick{\mathscr{C}_{0}} & \qw & \ctrl{3} & \qw & \qw & \qw & \ctrl{3} & \qw & \cdots &   & \qw & \qw &   \qw & \qw & \qw & \qw & \qw\\ 
&   &   &   &   &   &   &   &   &   &   & &  &   &  & &   \\
&   &   &   &   &   &   &   &   &   &   & &  &   &  &  &\\
\lstick{0} & \ustick{\ket{\psi}}  \qw & \multigate{2}{S} & \qw & \gate{U_{0}} & \qw & \multigate{2}{S^{-1}} & \qw & \cdots &   & \multigate{2}{S} & \qw &  \gate{U_{0}} & \qw & \multigate{2}{S^{-1}} & \qw & \qw & \ustick{Z \ket{\psi}}\\
& \vdots &   &  & \vdots &   & & \vdots & & \vdots &  & &  \vdots &   &  &  \vdots & &  \\
\lstick{N-1} & \qw & \ghost{S} & \qw & \gate{U_{N-1}} & \qw & \ghost{S^{-1}} & \qw & \cdots &   & \ghost{S} & \qw &  \gate{U_{N-1}} & \qw & \ghost{S^{-1}} & \qw & \qw \\
&   &   &   &   &   &   &   &   &   &   &   \\
}
\end{aligned}
\end{equation}
where $Z = U_{i_{(N)}} \dots U_{i_{(1)}}$ denotes the programmed disposition of unitaries, and
$\mathscr{C}_i$ ($i = 1,\dots, N$) denote a control register made of $n$ qubits. The circuit is just
the juxtaposition of $N$ copies of circuit (\ref{CnSWAPgenerico}), $\mathscr{C}_i$ being the control
system of the $i$the circuit. 

\begin{theorem} \label{circT1optimality}
Circuit (\ref{circT1}) is the most efficient implementation of Task \ref{t1}.
\end{theorem}

\Proof The proof is an immediate consequence of Theorem \ref{circT2optimality}, since the elements
of the sequence of unitaries $U_{i_{(N)}} \dots U_{i_{(1)}}$ are all independent, whence the circuit
that accomplishes Task \ref{t1} in the most efficient way must be a juxtaposition of $N$ optimal
circuits achieving Task \ref{t2}.\qed

\medskip
It is easy to see that, by construction, circuit (\ref{circT1}) is also the most efficient
implementation of Task \ref{t1} restricted to just the permutations of the unitaries, namely without repetitions.

\medskip In order to implement circuit (\ref{circT1}) we need $16N(N-1)$ \CNOT s and $18N(N-1)$
single-qubit operations, along with $N\log N$ control qubits plus $N-1$ ancillary qubits (only one
system qubit support the transformation).  In summary, the resource for Task \ref{t1} is
$\mathcal{O}(N^{2})$ \CNOT s and single-qubit operations and $\mathcal{O}(N^{2})$ qubits.

As an example, for $N = 2$, the circuit will be the following
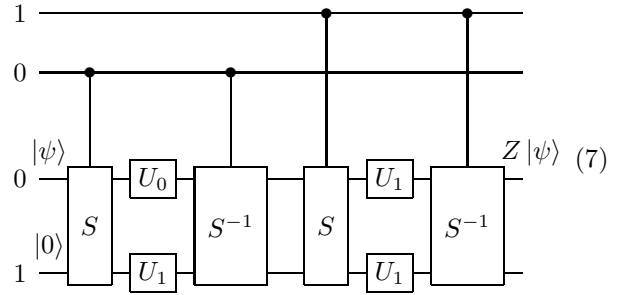
\begin{equation} \label{circT1N=2}
\begin{aligned}
\Qcircuit @C=.35em @R=1em { 
\lstick{1} & \qw & \qw&\qw & \qw & \qw & \qw & \qw & \qw & \qw & \qw   & \ctrl{5} & \qw & \qw & \qw & \ctrl{5} & \qw & \qw\\
&   &   &   &   &   &   &  &   & &   &   &   &   &   &   &    & \\
\lstick{0} & \qw & \qw&\ctrl{3} & \qw & \qw & \qw & \ctrl{3} & \qw & \qw & \qw  & \qw & \qw &   \qw & \qw & \qw & \qw & \qw\\ 
&   &   &   &   &   &   &   &   &   &   &   &   &  &   &  &   &   \\
&   &   &   &   &   &   &   &   &   &   &   &   &  &   &  &   &\\
\lstick{0} & \ustick{\ket{\psi}} \qw &\qw& \multigate{2}{S} & \qw & \gate{U_{0}} & \qw & \multigate{2}{S^{-1}} & \qw & \qw & \qw  & \multigate{2}{S} & \qw &  \gate{U_{1}} & \qw & \multigate{2}{S^{-1}} & \qw & \qw & \ustick{Z \ket{\psi}}\\
&   &   &   &  &   &   & &   & &   &  & &    &   &  &    & &  \\
\lstick{1} & \ustick{ \ket{0}} \qw & \qw &\ghost{S} & \qw & \gate{U_{1}} & \qw & \ghost{S^{-1}} & \qw &   \qw & \qw  & \ghost{S} & \qw &  \gate{U_{1}} & \qw & \ghost{S^{-1}} & \qw & \qw \\
&   &   &   &   &   &   &   &   &   &   &   &   \\
} 
\end{aligned}
\end{equation}
With the control registers prepared in a superposition state, the system qubit gets entangled with
the control register, e.~g for the control register in the state $\ket{\Phi_{1}}=\frac{1}{\sqrt2}[\ket{01}
+ \ket{10}]$ the output joint state of the system qubit supporting the input state $|\psi\>$ and the
control qubits will be $\ket{\Psi_{0}}=\frac{1}{\sqrt2}[ \ket{01} \otimes U_{0}\,U_{1}\ket{\psi} +
\ket{10} \otimes U_{1}\,U_{0}\ket{\psi}]$. Thus, if we program the control qubits in a tensor
product of $|+\>=\frac{1}{\sqrt2}[\ket{0}+\ket{1}]$ the output joint state will be entangled with
all possible dispositions of the unitaries of the set $\Uset$, a typical manifestation of quantum
parallelism.

\medskip We now investigate the possibility of programming all the possible permutations of $N$
unitary channels with a single use per channel, using quantum switches.  More precisely, the task is
the following:
\begin{task} \label{t3} Build a computational network with \QS s achieving any permutations of the
  unitaries of the set $\Uset$ (programmed on the state of a control register) with a single use per
  unitary.
\end{task}

A solution to Task \ref{t3} is provided by the quantum network with \QS s built with the following
procedure:
 
\begin{itemize}
\item[{\bf (N)}] Build the following input-output box-connections:
\begin{enumerate}
\item $U_0\rightarrow
  \map{S}_1^{(1)}\rightarrow\map{S}_1^{(2)}\rightarrow\ldots\rightarrow\map{S}_1^{(N-1)}\rightarrow U'_0$
\item for $k=1$ to $N-2$:
\begin{itemize}
\item[] $\map{S}_1^{(k)}\rightarrow\map{S}_2^{(k+1)}$, $\ldots\,$, $\map{S}_k^{(k)}\rightarrow\map{S}_{k+1}^{(k+1)}$, 
\end{itemize}
\item for $k=1$ to $N-1$:
\begin{itemize}
\item[] $U_k\rightarrow\map{S}_k^{(k)}\rightarrow\map{S}_{k-1}^{(k)}\rightarrow\ldots \rightarrow
\map{S}_2^{(k)}\rightarrow\map{S}_1^{(k)}$,
\end{itemize}
\item $\map{S}_1^{(N-1)}\rightarrow U'_1$, $\ldots\,$, $\map{S}_{N-1}^{(N-1)}\rightarrow
  U'_{N-1}$,
\end{enumerate}
\end{itemize}
Finally, to apply the sequence of unitary to the target qubit, one need to connect the output black
boxes $U'_i$ each other to obtain their product $U'_{N-1}U'_{N-2}\dots U'_{0}$.

\medskip
As an example, the computational networks for $N = 3,4$ are reported (the control registers omitted).
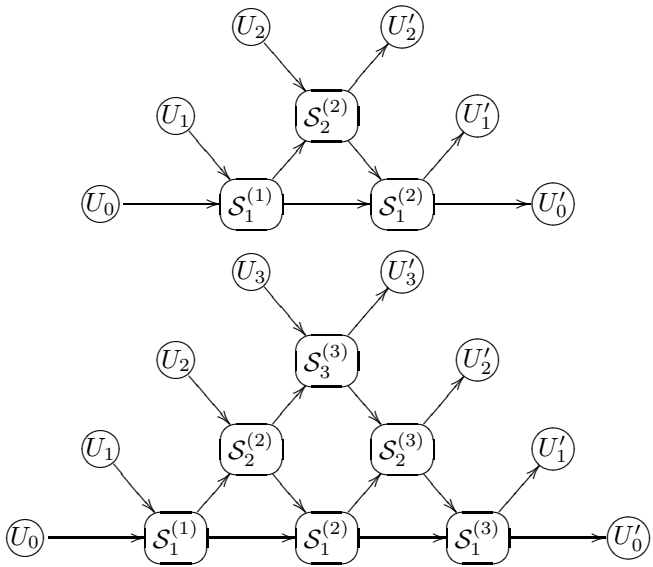
\begin{eqnarray*}
&
\xymatrix @C=.5em @R=1em @! {
& & *+[o][F-]{U_2} \ar[dr] & & *+[o][F-]{U'_2} &  \\
& *+[o][F-]{U_1} \ar[dr] & & *+[F-:<6pt>]{ \map S_2^{(2)} } \ar[dr] \ar[ur] & & *+[o][F-]{U'_1}   \\
*+[o][F-]{U_0} \ar[rr] & & *+[F-:<6pt>]{ \map S_1^{(1)} } \ar[ur] \ar[rr] & & *+[F-:<6pt>]{ \map S_1^{(2)} } \ar[rr] \ar[ur] &  & *+[o][F-]{U'_0}}
\nonumber\\
&\label{NETPERMTHREECHANNELS}\\
&
\xymatrix @C=.5em @R=1em @! {
& & & *+[o][F-]{U_3} \ar[dr] & & *+[o][F-]{U'_3} &  \\
& & *+[o][F-]{U_2} \ar[dr] & & *+[F-:<6pt>]{ \map S_3^{(3)} } \ar[dr] \ar[ur] & & *+[o][F-]{U'_2} &  \\
& *+[o][F-]{U_1} \ar[dr] & & *+[F-:<6pt>]{ \map S_2^{(2)} } \ar[dr] \ar[ur] & & *+[F-:<6pt>]{ \map S_2^{(3)} } \ar[dr] \ar[ur] & & *+[o][F-]{U'_1}   \\
*+[o][F-]{U_0} \ar[rr] & & *+[F-:<6pt>]{ \map S_1^{(1)} } \ar[ur] \ar[rr] & & *+[F-:<6pt>]{ \map S_1^{(2)} } \ar[rr] \ar[ur] & & *+[F-:<6pt>]{ \map S_1^{(3)} } \ar[rr] \ar[ur] & & *+[o][F-]{U'_0}}
\nonumber
\end{eqnarray*}
It is easy to see that the circuit uses overall $\sum_{n=1}^{N-1} n = \tfrac{1}{2} N (N-1)$ $\QS$s,
each with a single-qubit control register. 

\begin{table}[htdp]
\begin{center}
\begin{tabular}{|c|c|c|c|}
\hline 
$U'_2 U'_1 U'_0$ & \begin{tabular}{c} State of\\$s_{1}^{(1)}$  $s_{1}^{(2)}$  $s_{2}^{(2)}$ \end{tabular} &  $U'_2 U'_1 U'_0$ & \begin{tabular}{c} State of\\$s_{1}^{(1)}$  $s_{1}^{(2)}$  $s_{2}^{(2)}$ \end{tabular} \\
\hline
$U_{2}U_{1}U_{0}$ & \begin{tabular}{c}$\ket{0}\ket{0}\ket{0}$\\ $\ket{1}\ket{1}\ket{0}$\end{tabular}& $U_{2}U_{0}U_{1}$ & \begin{tabular}{c}$\ket{1}\ket{0}\ket{0}$ \\ $\ket{0}\ket{1}\ket{0}$ \end{tabular}\\
 \hline
 $U_{0}U_{2}U_{1}$ & $\ket{1}\ket{0}\ket{1}$ & $U_{1}U_{2}U_{0}$ & $\ket{0}\ket{0}\ket{1}$ \\
 \hline
$U_{1}U_{0}U_{2}$ & $\ket{0}\ket{1}\ket{1}$ & $U_{0}U_{1}U_{2}$ & $\ket{1}\ket{1}\ket{1}$ \\
 \hline
\end{tabular}
\caption{ The six possible outputs for the computational circuit (\ref{NETPERMTHREECHANNELS}) for
  $N=3$ and corresponding control state(s). The control qubit $s_i^{(j)}$ programs the \QS\ $\map
  S_i^{(j)}$\label{controlsgraphN=3}}
\end{center}
\end{table}

In Table \ref{controlsgraphN=3} we list the six possible outputs of the circuit for $N=3$ versus the
state of the control qubits $s_i^{(j)}$ (corresponding to the \QS s $\map S_i^{(j)}$).  Notice that
the same permutation can be achieved with different states of the control registers.

\begin{theorem} \label{circBEYONDoptimality} Network {\bf (N)} achieves Task \ref{t3} in the most
  efficient way, minimizing the number of $\QS$s and of ancillary control qubits.
\end{theorem}
\Proof The task is to set the ordering of $N$ different unitaries depending on the state of a
control register, using a network of $\QS$s. The permutation is defined by the relative ordering of
each pair of different unitaries, and there are $\frac{1}{2}N(N-1)$ of such pairs, which also equals
the number of $\QS$s and of their respective registers.\qed

\medskip In conclusion, we have seen that a quantum computation with programmable connections
between gates is more powerful than the usual quantum computation. This has been proved on the
specific task of programming all possible permutations of a cascade of unitary channels acting on a
qubit, where the number of uses per channels is dramatically reduced from $N$ to $1$.  For this task
the new {\em quantum switch} resource is needed, which can be programmed to switch the order of two
channels, each with a single use. A thorough analysis of the functions on channels that can be
physically achievable is in progress, and before its conclusion it is premature to state any
conjecture about universal functions for quantum computation with programmable connections.  In
particular, it is not clear yet whether the \QS\ function is universal or not. As an important
example, it can be proved that the $W$ operation considered in Ref. \cite{orebrucos} cannot be
achieved using only \QS s in addition to traditional quantum circuits. However, while \QS\ has a
straightforward operational interpretation, $W$ is an admissible mathematical map that presently
lacks physical interpretation. On practical grounds, we believe that the ultrafast switch of Ref.
\cite{kumarswitch} is a sufficient optical element for the implementation of the quantum switch,
being able to route entangled photons at high speed without disturbing the quantum state. This
practical possibility makes the implementation of \QS\ feasible in the near future, opening
important new options and perspectives in the design of new quantum algorithms.

\medskip We acknowledge Scott Aaronson for a stimulating discussion about Ref. \cite{beyondqc}.  G.
M. D. acknowledges interesting discussions with Joe Altepeter and Prem Kumar.  This work has been
supported by the Italian Ministry of Education through grant PRIN 2008, and by the EU through FP7 STREP
project COQUIT. Research at Perimeter Institute for Theoretical Physics is supported in part by
Canada through NSERC and by Ontario through MRI.

\end{document}

%% file: myQcircuit.tex
\usepackage[matrix,frame,arrow]{xy}
\usepackage{amsmath}

\newcommand{\ket}[1]{\left\vert{#1}\right\rangle}
\newcommand{\qw}[1][-1]{\ar @{-} [0,#1]}
\newcommand{\qwx}[1][-1]{\ar @{-} [#1,0]}

\newcommand{\gate}[1]{*{\xy *+<.6em>{#1};p\save+LU;+RU **\dir{-}\restore\save+RU;+RD **\dir{-}\restore\save+RD;+LD **\dir{-}\restore\POS+LD;+LU **\dir{-}\endxy} \qw}

\newcommand{\control}{*!<0em,.025em>-=-{\bullet}}

\newcommand{\ctrl}[1]{\control \qwx[#1] \qw}

\newcommand{\qswap}{*=<0em>{\times} \qw}
\newcommand{\multigate}[2]{*+<1em,.9em>{\hphantom{#2}} \qw \POS[0,0].[#1,0];p !C *{#2},p \save+LU;+RU **\dir{-}\restore\save+RU;+RD **\dir{-}\restore\save+RD;+LD **\dir{-}\restore\save+LD;+LU **\dir{-}\restore}
\newcommand{\ghost}[1]{*+<1em,.9em>{\hphantom{#1}} \qw}

\newcommand{\gategroup}[6]{\POS"#1,#2"."#3,#2"."#1,#4"."#3,#4"!C*+<#5>\frm{#6}}

\newcommand{\lstick}[1]{*!R!<.5em,0em>=<0em>{#1}}
\newcommand{\ustick}[1]{*!D!<0em,-.5em>=<0em>{#1}}

\newcommand{\Qcircuit}[1][0em]{\xymatrix @*=<#1>} 

